\documentstyle[12pt]{article}

\large
\newcommand{\be}{\begin{eqnarray}}
\newcommand{\ee}{\end{eqnarray}}

\begin{document}
\setlength{\baselineskip}{21pt}
\pagestyle{empty}
\vfill
\eject
\begin{flushright}
SUNY-NTG-93/2
\end{flushright}

\vskip 2.0cm
\centerline{\bf Spectral Density of the QCD Dirac Operator near Zero
Virtuality}
\vskip 2.0 cm
\centerline{J.J.M. Verbaarschot and I. Zahed}
\vskip .2cm
\centerline{Department of Physics}
\centerline{SUNY, Stony Brook, New York 11794}
\vskip 2cm

\centerline{\bf Abstract}
We investigate the spectral properties of a random matrix model, which
in the large $N$ limit, embodies the essentials of the QCD partition
function at low energy. The exact spectral density and its pair correlation
function are derived for an arbitrary number of flavors and zero topological
charge. Their microscopic limit provide the master formulae for
sum rules for the inverse powers of the eigenvalues of the QCD Dirac
operator as  recently discussed by Leutwyler and Smilga.
\vfill
\noindent
\begin{flushleft}
SUNY-NTG-93/2\\
February 1993
\end{flushleft}
\eject
\pagestyle{plain}

The issue of chiral symmetry is fundamental in QCD. Lattice simulations
indicate that the symmetry is spontaneously broken in the vacuum.
This fact has inspired a large body of work in an attempt to describe
the underlying mechanism. However, an explanation
from first principles is still elusive.
In a way, the
spontaneous breakdown of chiral symmetry reflects directly on the way
the quark states are delocalized in the vacuum
\cite{DIAKONOV-PETROV-1986,SHURYAK}, a situation reminiscent
of the delocalization of electrons in solids and the onset of conductivity
\cite{MCKANE-STONE-1981}.

A decade ago, Banks and Casher \cite{BANKS-CASHER-1980}
noted that the spontaneous breaking
of chiral symmetry is related to an accumulation in the quark spectral
density at zero virtuality.
In other words, as pointed out by Leutwyler and Smilga
\cite{LEUTWYLER-SMILGA-1992A}, the spacing of the
eigenvalues $\lambda$ is not proportional to $1/\sqrt{V_4}$
as in free space, but to $1/V_4$
($V_4$ is the space time volume). This
implies the existence of a microscopic limit of the spectral density
in which the thermodynamical
limit is taken for fixed values of $\lambda V_4$
\cite{SHURYAK-VERBAARSCHOT-1992F}.

In the extreme long wavelength limit, the QCD partition
function can be written as an integral over spacetime independent
Goldstone modes \cite{LEUTWYLER-SMILGA-1992A}. This implies
strong correlations between the eigenvalues of the Dirac operator
in the form of sum rules. These sum rules only
involve the microscopic limit of the spectral density.

We conjecture that the microscopic correlations between the eigenvalues
of the Dirac operator near zero virtuality are universal, and encoded in
the microscopic spectral density and its fluctuations. Our conjecture is
motivated by the following observations: First, the sum rules hold for
the massive Schwinger model \cite{SMILGA}.
Second, they are obeyed by the Dirac operator in a liquid of
instantons \cite{SHURYAK-VERBAARSCHOT-1992F} with correlations induced
by the fermion determinant \cite{SHURYAK}.
Third, a random matrix model
\cite{SHURYAK-VERBAARSCHOT-1992F} can be constructed that in the
the large $N$ limit reduces to the low energy limit of the
QCD partition function for a given value of the vacuum angle,
and thus satisfies all microscopic sum rules.
Fourth, in chaotic systems
microscopic fluctuations are universal and can be mapped on the invariant
random matrix ensembles
\cite{BOHIGAS-GIANNONI-SCHMIT-1984,MEHTA-1991}
\cite{SELIGMAN-VERBAARSCHOT-ZIRNBAUER-1984}.
Fifth, in the theory of mesoscopic systems, the Hofstadter model has a
spectral density that coincides with that of the above random matrix model
in the quenched  approximation \cite{HOFSTADTER-1976}.

The above observations suggest that a random matrix model with the general
symmetries of the QCD partition, might be key in understanding the general
implications of chiral symmetry breaking in the QCD vacuum.
The aim of the present paper is to construct the microscopic limit of its
spectral density and correlations thereof. They constitute the master equations
for the Leutwyler-Smilga spectral sum rules.
We restrict ourselves to the simplest possible
case of  zero total topological charge. The model
is outlined in the next section. Using the orthogonal polynomial approach
to random matrix theory, we derive an explicit expression for the
spectral density in the chiral limit for an arbitrary number of flavors
and zero topological charge.

Consider a system of $n = N/2$ zero
modes and $n$ anti-zero modes with interaction given by the
$n\times n$ overlap matrix $T$. For $N_f$ flavors, the partition function
that reflects the chiral structure of QCD is given by
\cite{NOWAK-VERBAARSCHOT-ZAHED-1989A,SIMONOV-1991,SHURYAK-VERBAARSCHOT-1992F}
\be
Z = \int {\cal D}T P(T)\prod_f^{N_f}\det \left (
\begin{array}{cc} m_f & iT\\
                 iT^\dagger & m_f
\end{array} \right ),
\ee
where the integral is over the real and imaginary parts of the matrix elements
of the arbitrary complex matrix $T$ $i.e.$, ${\cal D}T$ is the Haar measure.
In agreement with the maximum entropy principle \cite{BALIAN-1968}
the distribution function of the overlap matrix
elements $P(T)$ is chosen Gaussian
\be
P(T) = \exp(-\frac N{2\sigma^2} T T^{\dagger}).
\ee
The symplectic structure is a manifest consequence of chiral symmetry,
and implies that the quark eigenvalues occur in pairs. The density of the
zero modes, $N/V_4$, is taken equal to 1, which allows us to identify
the space-time volume and the number of zero modes.

The order parameter in the study of the spontaneous breaking of chiral
symmetry is the quark condensate defined through
\be
\Sigma_f = <\bar q_f q_f>= \lim_{m_f\rightarrow 0}
\lim_{N\rightarrow \infty}-\frac {1}N \frac d{d m_f} \log Z,
\ee
where the chiral limit is to be taken $after$ the thermodynamical limit.
By writing the determinant as the product
$\prod(\lambda^2_n + m^2_f)$
one obtains the Banks-Casher formula \cite{BANKS-CASHER-1980}
\be
\Sigma = \pi<\rho_C(0)>_{Z},
\ee
where the average $<\cdots>_{Z}$ is with respect to
the partition function (1).
The continuum spectral density is defined by
\be
\rho_C(\lambda) =\lim_{{\rm all} \,\,m_f\downarrow 0}
 \lim_{N \rightarrow \infty}\frac 1N \rho(\lambda),
\ee
and the spectral density $\rho(\lambda)$
is
\be
\rho(\lambda) = \sum \delta(\lambda -\lambda_n),
\ee
where the eigenvalues $\pm \lambda_n$ are the nonzero eigenvalues of the
overlap
matrix in the chiral limit.
As has been shown in
\cite{SHURYAK-VERBAARSCHOT-1992F} the parameter $\sigma$ can be identified as
$\sigma = 1/\Sigma$.

The sum rules recently discussed by Leutwyler and Smilga
\cite{LEUTWYLER-SMILGA-1992A} for the QCD Dirac operator using
chiral perturbation theory involve the average of the sums
\be
\sum_n \frac 1{N^{2p}\lambda_n^{2p}},\qquad
\sum_{nm} \frac 1{N^{p+q}\lambda_n^{p}\lambda_m^{q}}\ee
which can be rewritten as
\be
\int_0^\infty dx \frac {\rho_S(x)}{x^{2p}}, \qquad
\int_0^\infty dx dy \frac {\rho_S(x) \rho_S(y)}{x^{p} y^q},
\ee
respectively,
where we have introduced the microscopic spectral density defined by
\be
\rho_S(x) = \lim_{N\rightarrow \infty} \frac 1N \rho(\frac xN).
\ee
In this letter our primary interest is to derive analytical
expressions for the average of $\rho_S(x)$ and its pair correlation function
that summarize the sum rules.

The partition function (1) can be evaluated by rewriting the matrix
integration in polar coordinates. For an arbitrary complex matrix we
may write \cite{HUA-1963}
\be
T = U \Lambda V^{-1},
\ee
where $U$ and $V$ are unitary matrices and $\Lambda$ is a positive definite
diagonal matrix. Since the r.h.s has $N$ more degrees of freedom
than the l.h.s., one has to impose constraints on the new integration
variables.
This can be achieved \cite{HUA-1963}
by restricting $U$ to the coset $U(N)/U(1)^{N}$, where
$U(1)^{N}$ is the diagonal subgroup of $U(N)$. The Jacobian of this
transformation, that  depends only on the eigenvalues $\lambda_k$ of $\Lambda$,
is given by
\be
J(\Lambda) = \prod_{k < l}(\lambda^2_k -\lambda^2_l)^2 \prod_k \lambda_k.
\ee
The integrations over the eigenvalues and the unitary matrices decouple. The
latter only result in an overall irrelevant constant factor and can be ignored.
The eigenvalue distribution is thus given by
\be
\rho_n(\lambda_1, \cdots, \lambda_n) = J(\Lambda)
\prod_{f}\prod_k(\lambda_k^2 +m_f^2) \exp(-\frac n{\sigma^2}\sum_{k=1}^n
\lambda_k^2).
\ee
The spectral density $\rho_1(\lambda)$ is obtained by integration over
the remaining $n-1$ eigenvalues
\be
\rho_1 (\lambda_1) = \int \prod_{k=2}^n d\lambda_k
\rho_n(\lambda_1, \cdots, \lambda_n).
\ee
These integrals can be evaluated with the help of the methods developed
by Dyson, Mehta and Wigner (see \cite{MEHTA-1991} for references).
The main ingredient is to write the product over the differences of
the eigenvalues as a Vandermonde determinant, $i.e.$
\be
\prod_{k < l}(\lambda^2_k -\lambda^2_l)^2 =
    \left | \begin{array}{ccc} 1     & \cdots  &  1     \\
                               \cdot &         & \cdot  \\
                               \cdot &         & \cdot  \\
                              \lambda_1^{2(n-1)} &\cdots &\lambda_n^{2(n-1)}
    \end{array}    \right|^2.
\ee
which up to a constant can be rewritten in terms of orthogonal polynomials
$P_k$ as
\be
    \left | \begin{array}{ccc} P_0(\lambda_1^2) & \cdots  & P_0(\lambda_n^2)\\
                               \cdot &         & \cdot  \\
                               \cdot &         & \cdot  \\
                       P_{n-1}(\lambda_1^2) &\cdots  &P_{n-1}(\lambda_n^2)
    \end{array}    \right|.
\ee
The $P_k$ will be chosen orthogonal according to the weight function
\be
\int_0^\infty d(\lambda^2) (\lambda^2 + m^2)^{N_f} \exp(-\frac n{\sigma^2}
\lambda^2) P_k(\lambda^2) P_l(\lambda^2) = \delta_{kl}.
\ee
For $m = 0$ these
polynomials are well known,
\be
P_k(s) = \left ( \frac n{\sigma^2}
\frac{k!}{\Gamma( N_f +k +1)}\right )^{1/2} L_k^{N_f}(\frac {sn}{\sigma^2}),
\ee
where the $L_k^{N_f}$ are the associated Laguerre (Sonine) polynomials.

The determinants can be expanded according to Cramers rule. All integrals
can be performed immediately by orthogonality and, up to an overall constant,
 we are left with
\be
\rho_1(\lambda) = \sum_{k=0}^{n-1}
\frac{k!}{\Gamma( N_f +k +1)} L_k^{N_f}(z)
L_k^{N_f}(z) z^{N_f +1/2} \exp( -z),
\ee
where $z$ is defined by
\be
z = \frac {n\lambda^2}{\sigma^2}
\ee
The sum can be evaluated exactly with the Christoffel-Darboux formula,
resulting in
\be
\rho_1(\lambda) = \frac {n}{\sigma^2}
\frac{n!}{\Gamma( N_f +n)}\left ( L_{n-1}^{N_f}(z)L_{n-1}^{N_f+1}(z)-
L_{n}^{N_f}(z)L_{n-2}^{N_f+1}(z)\right )
z^{N_f +1/2} \exp( -z),
\ee
which, up to a normalization constant,
constitutes the exact spectral density of the model (1).
The microscopic limit is obtained by taking $N \rightarrow \infty$ while
keeping $N\lambda = x$ fixed (remember that $n = N/2$).
This can be achieved from the asymptotic relation
\be
\lim_{n \rightarrow \infty} \frac 1{n^\alpha} L_n^\alpha(\frac xn) =
x^{-\frac {\alpha}{2}} J_\alpha(2 \sqrt x),
\ee
where $J_\alpha$ is the ordinary Bessel function of degree $\alpha$.
The result for the microscopic spectral density is
\be
\rho_S(x)  = \frac {\Sigma^2 x}{2} (J^2_{N_f}(\Sigma x) -J_{N_f+1}(\Sigma x)
             J_{N_f-1}(\Sigma x)).
\ee
The normalization constant follows from the asymptotic behaviour of the
Bessel function and the Banks-Casher relation.

This formula reproduces all diagonal sum rules of Leutwyler and Smilga. $e.g.$,
the sum
\be
\sum_n \frac 1{N^{2p} \lambda_n^{2p}}
\ee
can be converted into an integral over the microscopic variable $x =\lambda N$
resulting in
\be
\int_0^\infty \frac {\rho_S(x) dx}{x^{2p}} =
\left(\frac {\Sigma}{2}\right)^{2p}\,\frac{\Gamma (2p-1)\Gamma (N_f-p+1)}
{\Gamma (p)\Gamma (p+1) \Gamma (N_f + p )}
\ee
The above spectral density thus summarizes all sum rules and yields
new sum rules, e.g. sum rules for noninteger values of $p$.
The spectral density for 0, 1 and 2 flavors are shown by the full, dashed
and dotted curves in Fig. 1.

The non-diagonal sum rules require correlations in the spectral density.
These correlators can be calculated using similar techniques. Also, the
above arguments can be extended to finite quark masses
\cite{VERBAARSCHOT-ZAHED-1993}.

The two point correlation function $\rho_2 (\lambda ,\lambda')$
follows from (12) by integration over $n-2$ eigenvalues and subtracting
the disconnected part. Using again the properties of the orthogonal
polynomial and the Christoffel-Darboux formulae we have

\be
\rho_2 (\lambda, \lambda') = \left(
\frac{n}{\sigma^2}\frac{n!}{\Gamma(n+N_F)}\right)^2\,\,(zz')^{N_F+1/2}\,
e^{-(z+z')}
\left(\frac{L_n^{N_F}(z)L_{n-1}^{N_F}(z')-L_n^{N_F}(z')L_{n-1}^{N_F}(z)}
{z-z'}\right)^2
\label{two}
\ee
with $z=n\lambda^2/\sigma^2$ and $z'=n\lambda'^2/\sigma^2$.
The microscopic limit $\rho (x, y) ={\rm lim}_{N\rightarrow\infty}
\rho_2 (x/N ,y/N)$ is obtained by using similar arguments to
the ones used for the spectral density. The result is

\be
\rho (x, y) = \Sigma^2\,xy\,\left(
\frac{xJ_{N_F} (\Sigma x)J_{N_F-1} (\Sigma y)-
      y J_{N_F}(\Sigma y)J_{N_F-1} (\Sigma x)}{x^2-y^2}\right)^2
\label{twomicro}
\ee
The microscopic limit (\ref{twomicro}) can be used to derive off-diagonal
sum rules for the QCD Dirac operator. Indeed,

\be
<\left(\sum_n\frac 1{N^2\lambda_n^2}\right)^2>-<\sum \frac 1{N^4\lambda_n^4}>
-<\sum_n\frac 1{N^2\lambda_n^2}>^2 =
\int dx dy \frac{\rho (x, y)}{ x^2y^2}
\ee
We were not able to evaluate this integral analytically, but numerically
the result is given by
\be
\frac{\Sigma^4}{16 N_f^2 (N_f + 1)},
\ee
\noindent for $N_f = 1, 2, \dots, 20$ with an accuracy of better than
1 part in $10^5$ and agrees completely with Leutwyler and
Smilga \cite{LEUTWYLER-SMILGA-1992A}.

We have shown that the microscopic spectral density following from
the Gaussian chiral ensemble of complex matrices reproduces all
microscopic sum rules for the QCD Dirac operator. The corresponding
random matrix model reflects solely on chiral symmetry.
We have strong indications that
the microscopic spectral density is universal and should
be a solid property of the QCD Dirac operator near zero virtuality. The overall
spectral density is not. It would be interesting to see how the former
compares to the true QCD spectral density following from lattice
simulations. Our approach is relevant for discussing similar issues
related to QCD in two and three dimensions, as well as the spectral
properties of strongly coupled QED.
Finally, we note that the spectral density for zero flavors describes the
spectral correlations in the Hofstadter model as used in the framework of
universal conductance fluctuations
\cite{HOFSTADTER-1976,ANDO-1991,SLEVIN-NAGAO-1993}.
The physical analogy is striking,
emphasizing once more the universality of the bulk structure of chiral
symmetry breaking in the QCD vacuum.
\vglue 0.6cm
{\bf \noindent  Acknowledgements \hfil}
\vglue 0.4cm
 The reported work was partially supported by the US DOE grant
DE-FG-88ER40388.
We would like to thank A. Smilga and M.A. Nowak for useful discussions.

\vfill
\eject
\newpage
\addtocounter{page}{1}
\noindent
{\bf Figure Caption}
\vskip 0.5cm
\noindent
Fig. 1. The microscopic spectral density $\rho_S(x)$ as a function of
the microscopic variable $x$ for 0 (full line), 1 (dashed line) and
2 (dotted line) flavors.

\vfill
\eject
\addtocounter{page}{-2}
\newpage
\setlength{\baselineskip}{15pt}

\bibliographystyle{aip}

\end{document}